# Why Anticipatory Sensing Matters in Commercial ACC Systems under Cut-In Scenarios: A Perspective from Stochastic Safety Analysis

Hao Zhang[a], Sixu Li[a], Zihao Li[a], Mohammad Anis[a], Dominique Lord[a], Yang Zhou[a,*]

[a]Zachry Department of Civil & Environmental Engineering, Texas A&M University, College Station, TX 77843 United States

**Abstract**

This study presents an analytical solution for the vehicle state evolution of Adaptive Cruise Control (ACC) systems under cut-in scenarios, incorporating sensing delays and anticipation using the Lambert W function. The theoretical analysis demonstrates that the vehicle state evolution and the corresponding safety of ACC in cut-in situations are influenced by multiple factors, including the original leading vehicle's state, the initial conditions of the cut-in vehicle, subsequent cut-in maneuvers, sensing delays, and the ACC's anticipation capabilities.

To quantitatively assess these influences, a series of numerical experiments were conducted to perform a stochastic safety analysis of ACC systems, accounting for embedded sensing delays and anticipation, using empirically calibrated control parameters from real-world data. The experiments revealed that the impact of sensing delays on ACC is multifaceted. Specifically, sensing delays negatively affect ACC stability, with the severity increasing as the delay lengthens. Furthermore, collision risk in cut-in scenarios becomes more significant with sensing delays, particularly when the cut-in vehicle is slower than the following vehicle and when cut-ins are aggressive.

However, anticipation plays a crucial role in mitigating these risks. Even with a 0.6-second anticipation, collision risk can be reduced by 91% in highly adverse scenarios. Finally, both sensing delays and anticipation have effects that intensify with their duration. An anticipation period of 2 seconds effectively ensures safety in aggressive cut-in conditions, even in the presence of sensing delays.

## 1. Introduction

Adaptive cruise control (ACC), serving as the basic driver-assistance automation function (i.e., Automation Level I according to SAE standard) (Taxonomy, 2016), is designed to automatically adjust a vehicle's acceleration to maintain a safe following distance from the immediately leading vehicle (Milanés and Shladover, 2014). While ACC has been extensively studied in terms of safety and stability through theoretical and simulation-based research (Bouadi et al., 2024; Xiao and Gao, 2010; Zhou et al., 2017), the safety benefits of commercially available ACC systems under cut-in conditions remain unclear.

Compared to the pure car-following process, vehicle cut-in scenario is influenced by more factors (Lu et al., 2023; Milanés and Shladover, 2016). The vehicle cut-in scenario introduces disturbances to the ACC system by causing deviations from the target spacing of the following vehicle and the cut-in vehicle's maneuvers. These disturbances will propagate through traffic flow and further render safety risks. Vehicle cut-ins are safety-critical events as they cause abrupt changes in the gaps between the following and cut-in vehicles. Empirical studies have shown that vehicle cut-ins are a major cause of traffic crashes and congestion (Liu et al., 2021). Additionally, cut-ins present significant challenges to advanced driver assistance systems (ADAS) and automated vehicles, often triggering emergency braking, which can increase fuel consumption and emissions.

Thus, it is significant to understand the impact of delay and anticipation on ACC safety under cut-in scenarios. A series of experiments by Zhao et al. (2020) demonstrated that relative distance, relative speed, and the speed of the cut-in vehicle significantly impact the comfort of surrounding drivers. Wang et al., (2019) analysed cut-in behaviours from naturalistic driving data, showing that cut-ins typically have shorter time-to-collision values compared to other lane changes. Fu et al., (2019) proposed a human-like car-following model considering cut-in scenarios, improving deriving safety and comfort. These studies offer valuable insights into understanding cut-in behaviours; however, they primarily rely on numerical experiments and data analysis (Lee et al., 2004). Z. Li et al. (2024b) proposed a theoretical framework to model ACC state evolution under cut-in scenarios, providing a comprehensive understanding of the impact of cut-in conditions on ACC safety and stability. However, the approach largely ignores the sensing delay as well as the actuation lags which are key features analysed by a wide range of studies (Bouadi et al., 2024; Hu et al., 2019; Wang et al., 2018). This treatment provides over-optimistic results, and large ignore the physical process that the original leading vehicle may also exert the impact on the ACC behaviour during cut-in process caused by the sensing delay.

While important, the prevailing researches on the sensing delay adverse impact on the traffic largely focus on the pure car following process and stability analysis (Wang et al., 2018; Zhang and Orosz, 2016). Some studies developed longitudinal dynamic and control models of ACC systems considering sensing delay and actuator lag (Davis, 2013a; Orosz et al., 2010). Based on these delay-embedded ACC control strategies, the influence of delays on the local and string stability of ACC systems and the stability conditions are discussed (Davis, 2013b, 2012). For example, the string stability region of the ACC platoon considering sensing delay and actuator lag was derived, and their impact on string stability was also investigated (Hu et al., 2019; Khound et al., 2022). However, stability and safety, while interrelated, are not equivalent, especially under disturbance conditions (S. Li et al., 2024; Li, 2022; Z. Li et al., 2024). Stability describes a vehicle's ability to remain near an equilibrium state under disturbance (local stability) or maintain dampening or constant disturbances through a vehicle platoon (string stability) (Zhou et al., 2019), which does not guarantee collision avoidance or prevent traffic voids during this process. For example, theoretical frameworks and numerical experiments proposed by Z. Li et al. (2024) have demonstrated that cut-in manoeuvres can lead to traffic oscillations and rear-end collisions. Though advancing, sensing delay, one of the most important ACC characteristics has been ignored.

As a compensation of sensing delay, anticipation is widely observed for both human driven vehicles and adopted in the design of automated vehicles. As observed by Zheng et al. (2013), the immediate human driven following vehicles usually exhibit the anticipatory behaviour during the lane-changing process. Similarly, recent research developed anticipatory control algorithms for immediate follower which are automated vehicles. For example, Kamal et al., (2022) developed a control strategy incorporating look-ahead anticipation, which improved driving comfort during high-speed driving, though the potential safety benefits were not examined. Shi et al. (2024) introduced a predictive deep reinforcement learning approach for longitudinal control to enhance the safety of connected automated vehicles (CAVs). Additionally, some studies have employed model predictive control (MPC) to compensate for sensing delays and enhance safety of automated vehicles (Nahidi et al., 2019; Sun et al., 2022; Wang et al., 2018). Thus, it is crucial to highlight the benefits of anticipatory sensing for ACC, especially in safety-critical cut-in scenarios.

To address these gaps, this study utilizes delay differential equation (DDE) to model the ACC with sensing delay and anticipation under cut-in scenario. A general solution of the DDE is developed based on the Lambert W function, which not only provides a comprehensive theoretical framework for analyzing the state evolution of the ACC system but also offers key insights into the system's behavior under different control parameters.

Additionally, a stochastic behaviour embedded high-fidelitous surrogate safety measure (SSM) is developed to unveil the benefits of the commercial ACC under cut-in scenario incorporating sensing delay and anticipation (S. Li et al., 2024; Z. Li et al., 2024). The inclusion of stochastic behavior in the SSM enables a more realistic and robust analysis by accounting for the inherent uncertainties in real-world driving. Numerical experiments, based on empirically calibrated control parameters derived from commercial ACC datasets, are conducted to validate the theoretical derivation (Zhou et al., 2022). These experiments provide quantitative insights into the effects of control parameters, cut-in vehicle behavior, and the original leading vehicle on the the safety performance of ACC system.

This paper is organized as follows. Section 2 develops the theoretical derivation of ACC system considering sensing delays and anticipation under cut-in scenarios and describes the numerical experiments setting (page 3). Section 3 describes and discuss the results of the scenarios-based experiments and sensitivity analyses (page 8). Section 4 concludes this study (page 16).

## 2. Methodology and Derivation

This section describes the vehicles' motion in cut-in scenarios considering sensing delay and anticipation in a form of DDE. Three vehicles are considered in the analysis: the cut-in vehicle, the following vehicle, and the original leading vehicle. To begin with, we follow the Society of Automotive Engineers (SAE) standard, the constant time gap policy is utilized for the ACC car-following (Zhou et al., 2017). The equilibrium spacing can be defined as follows:

$$s_f^*(t) = v_f(t) \times \tau^* + l \tag{1}$$

where $s_f^*(t)$ is the equilibrium spacing of the following vehicle at time $t$. $\tau^*$ is the desired constant time gap, $l$ is the standstill distance, and $v_f(t)$ is the velocity of following vehicle. The actuation lag of the following vehicle is characterized by $T_L$, where the realized acceleration rate $a_f(t)$ of the following vehicle is given by the first-order approximation (Nagatani and Nakanishi, 1998):

$$\dot{a}_f(t) = -\frac{1}{T_L}a_f(t) + \frac{1}{T_L}u_f(t) \tag{2}$$

where $u_f(t)$ is the demand acceleration rate. As shown in Fig. 1, caused by the sensing delay, there are two leader-follower pairs in the cut-in scenario at different time points. Specifically, when the vehicle cut-in occurs ($t = 0$), the cut-in vehicle becomes the new leading vehicle for the following vehicle. However, due to the sensing delay $\theta$, at $t = 0$, the following vehicle receives the information from the original leading vehicle. Thus, when $0 \leq t \leq \theta$, the following vehicle is following the original leading vehicle. When $t > \theta$, the following vehicle begins to follow the cut-in vehicle. The deviations from equilibrium spacing are defined as $\Delta s_l(t) = s_l(t) - s_f^*(t)$ and $\Delta s_c(t) = s_c(t) - s_f^*(t)$, representing the spacing deviation of two leader-follower pairs. $s_l(t)$ denotes the spacing between the following vehicle and original leading vehicle, and $s_c(t)$ is the spacing between the following vehicle and its new leading vehicle, the cut-in vehicle, which are calculated as: $s_l(t) = p_l(t) - p_f(t)$ and $s_c(t) = p_c(t) - p_f(t)$. $p_l(t)$ and $p_c(t)$ denote the position of the original leading vehicle and cut-in vehicle. The corresponding velocity differences are $\Delta v_l(t) = v_l(t) - v_f(t)$ and $\Delta v_c(t) = v_c(t) - v_f(t)$, where $v_l(t)$, $v_c(t)$, and $v_f(t)$ are the velocities of the original leading vehicle, cut-in vehicle, and following vehicle.

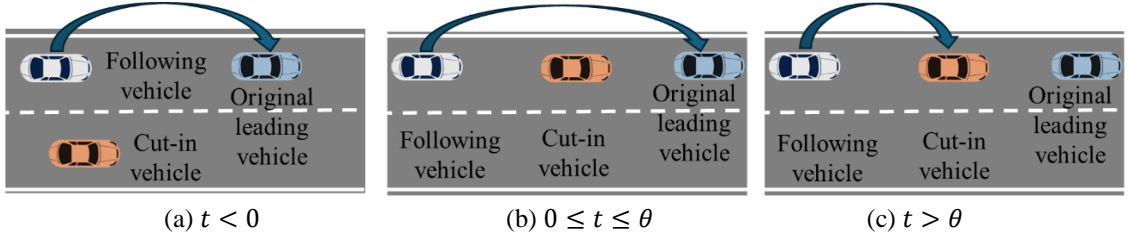

(a) $t < 0$  (b) $0 \leq t \leq \theta$  (c) $t > \theta$

Fig. 1. Illustration of the car-following pairs of the ACC cut-in scenario considering sensing delay ($\theta$).

By defining the state between the follower and original leading vehicle as $x_l(t) = [\Delta s_l(t), \Delta v_l(t), a_f(t)]^T$, the state between the follower and cut-in vehicle as $x_c(t) = [\Delta s_c(t), \Delta v_c(t), a_f(t)]^T$, the system considering actuation lag and sensing delay can be reformulated as a linear DDE:

$$\dot{x}_c(t) = \begin{cases} Ax_c(t) + BKx_l(t-\theta) + Da_c(t), & 0 \leq t < \theta \\ Ax_c(t) + BKx_c(t-\theta) + Da_c(t), & t \geq \theta \end{cases} \tag{3}$$

where $A = \begin{bmatrix} 0 & 1 & -\tau^* \\ 0 & 0 & -1 \\ 0 & 0 & -\frac{1}{T_L} \end{bmatrix}$, $B = \begin{bmatrix} 0 \\ 0 \\ \frac{1}{T_L} \end{bmatrix}$, $D = \begin{bmatrix} 0 \\ 1 \\ 0 \end{bmatrix}$, $a_c(t)$ denotes the acceleration rate of the cut-in vehicle. $Kx_c(t-\theta)$ represents the linear feedback control input with sensing delay $\theta$. $K = [k_s, k_v, k_a]$ are the feedback gains for the deviation from equilibrium spacing ($k_s$), speed difference ($k_v$), and acceleration ($k_a$), respectively. $x_l(t)$ can be calculated by solving the following DDE:

$$\dot{x}_l(t) = Ax_l(t) + BKx_l(t-\theta) + Da_l(t), -\theta \leq t < 0 \tag{4}$$

where $a_l(t)$ represents the acceleration rate of the original leading vehicle. The anticipatory sensing, denoted as $\varphi$, can be involved by advancing the moment when the follower vehicle reacts to the cut-in vehicle. When $\varphi > \theta$, the following vehicle responses to the vehicle cut-in before it occurs. Thus, the state evolution horizon in Eq. (3) should be extended to $t = t^l$ and $t^l$ is a negative value. Then the equation involves both the sensing delay and anticipation can be defined as:

$$\dot{x}_c(t) = \begin{cases} Ax_c(t) + BKx_l(t-\theta) + Da_c(t), & t^l \leq t < \theta - \varphi \\ Ax_c(t) + BKx_c(t-\theta) + Da_c(t), & t \geq \theta - \varphi \end{cases} \tag{5}$$

The state evolution of the follower and original leading vehicle is revised as follows:

$$\dot{x}_l(t) = Ax_l(t) + BKx_l(t-\theta) + Da_l(t), t^l < t < -\varphi \tag{6}$$

Taking Eq. (3) as an example, an analytic solution of the DDE of based on the concept of the Lambert W function (Asl and Ulsoy, 2003; Yi and Ulsoy, 2006) is given as:

$$x_c(t) = \underbrace{\sum_{k=-\infty}^{+\infty} e^{S_k t} C_k^I}_{\text{free response}} + \underbrace{\int_0^t \sum_{k=-\infty}^{+\infty} e^{S_k(t-\varepsilon)} C_k^N Da_c(\varepsilon) \, d\varepsilon}_{\text{forced response}} \tag{7}$$

The main advantage of this analytical solution for the DDE is that it has a similar form to the general solution of the ordinary differential equation (ODE). Specifically, the $S_k$, a $3 \times 3$ matrix, denotes the inherent characteristics of the system, independent of initial conditions and input. The eigenvalues of $S_k$ determine the stability of the linear delay system. $S_k$ is calculated as:

$$S_k = \frac{1}{\theta} W_k(BK\theta Q_k) + A \tag{8}$$

where $W_k$ denotes the $k$ branch of matrix Lambert W function, $Q_k$ is a $3 \times 3$ matrix. For each branch $k$, the eigenvalues $\hat{\lambda}_{ki}$, $i = 1,2,3$ of matrix $BK\theta Q_k$ and the corresponding eigenvector matrix $V_k$ is calculated. Then $W_k(BK\theta Q_k)$ can be calculated as:

$$W_k(BK\theta Q_k) = V_k \begin{bmatrix} w_k(\hat{\lambda}_{k1}) & 0 & 0 \\ 0 & w_k(\hat{\lambda}_{k2}) & 0 \\ 0 & 0 & w_k(\hat{\lambda}_{k3}) \end{bmatrix} V_k^{-1} \tag{9}$$

where $w_k$ denotes the scalar Lambert W function, which is defined as the $k$ branch solution of the converse relation of the function $y = f(w) = we^w$ and it can be formulated as:

$$w_k(y)e^{w_k(y)} = y \tag{10}$$

Because the Eq. (10) has multiple solutions, the Lambert W function has infinite branches. Fig. 2 demonstrates the different solution branches of the Lambert W function; different branches are described in different colors and the real and imaginary parts of the solution are described in solid and dotted lines, respectively. The red dotted line denotes the minimum value of the function $f(w) = we^w$. As demonstrated by Yi and Ulsoy (2006), with more branches, the results show better agreement with the true solution. Due to the branches of the Lambert W function, the eigenspectrum of $S_k$ is infinite and the rightmost (largest real parts) eigenvalues determines the system stability. As proved by Asl and Ulsoy (2003), when $BK$ and $A$ commute ($BKA = ABK$), principal branch ($k = 0$) always determines the system stability. However, in this study, it is validated that the two matrices do not commute. For this case, a reasonable conjecture given by Yi et al. (2007) suggests that the stability can be determined by the branches $k = 0$ or $k = \pm 1$. The system is stable when the values of the real parts of the rightmost eigenvalues are negative. To calculate $S_k$, $Q_k$ should first be solved as:

$$W_k(BK\theta Q_k)e^{W_k(BK\theta Q_k)+A\theta} = BK\theta \tag{11}$$

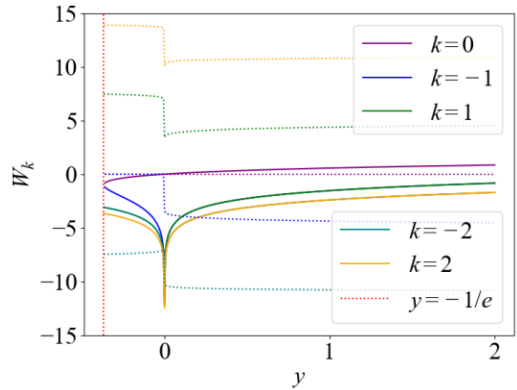

Fig. 2. Illustration of the branches of Lambert W function solution.

Based on Eq. (11), $Q_k$ is usually obtained numerically utilizing the 'fslove' function in MATLAB (Asl and Ulsoy, 2003). The free response's coefficient $C_k^I$ is a function of both the initial state between the follower and cut-in vehicle ($x_c(0)$) and the state between the follower and original leading vehicle ($x_l(t)$), which can be approximated as follows:

$$\begin{Bmatrix} x_c(0) \\ x_l\left(\frac{-\theta}{2N}\right) \\ x_l\left(\frac{-2\theta}{2N}\right) \\ \vdots \\ x_l(-\theta) \end{Bmatrix} = \begin{bmatrix} e^{S_{-N}0} & \cdots & e^{S_N 0} \\ e^{S_{-N}\frac{-\theta}{2N}} & \cdots & e^{S_N\frac{-\theta}{2N}} \\ e^{S_{-N}\frac{-2\theta}{2N}} & \cdots & e^{S_N\frac{-2\theta}{2N}} \\ \vdots & \vdots & \vdots \\ e^{S_{-N}-\theta} & \cdots & e^{S_N-\theta} \end{bmatrix} \begin{Bmatrix} C_{-N}^I \\ C_{-(N-1)}^I \\ C_{-(N-2)}^I \\ \vdots \\ C_N^I \end{Bmatrix} \qquad (12)$$

where $N$ represents the branches of the Lamber W function and $N \to \infty$ enables the $C_k^I$ approximate the accurate value infinitely. Based on the free response of the general solution, two conclusions can be derived: 1) due to sensing delay, the initial state of the ACC under cut-in scenario involves not only the initial state of the cut-in vehicle but also the original leading vehicle's state before cut-in. 2) The inherent characteristic, e.g., stability, of the ACC is influenced by the sensing delay. The forced response of the Eq. (3) depicts the vehicle behaviors contributed by cut-in vehicle's acceleration and deceleration after cut-in ($a_c(t)$). $C_k^N$ is dependent on $A$, $BK$, and sensing delay $\theta$. The coefficients $C_k^I$ and $C_k^N$ are usually solved by numerical methods. More details about the analytical solution of the DDE can be found in the referenced studies (Asl and Ulsoy, 2003; Yi and Ulsoy, 2006). Based on the forced response, the cut-in vehicle's acceleration and braking behaviors' impact on the ACC is influenced by the sensing delay.

For the ACC car-following control, the control input ($u_f(t)$) represents the demand deceleration rate of the following vehicle, which has a boundary value defined as $u_f^b$. However, it is complex to theoretically involve the boundary cases in the general solution of the DDE. Because this study focuses on the sensing delay and anticipation's impact on the ACC safety, the safety critical situation is utilized to simplify the problem. Specifically, this study assumes the following vehicle decelerates at the minimum deceleration rate ($u_f^b$) after the cut-in vehicle is sensed by the following vehicle at time $\theta$. In this case, if the vehicle collision is detected, the vehicle collision will also exist in real world because the cut-in vehicle decelerates at the minimum deceleration rate cannot avert the collision. If the vehicle collision is not detected, the vehicle collision can also be avoided in real world. Thus, in terms of the collision analysis, it is reasonable to assume the following vehicle decelerates at the maximum deceleration rate considering the worst case. The Eq. (5) can be revised as:

$$\dot{x}_c(t) = \begin{cases} Ax_c(t) + BKx_l(t-\theta) + Da_c(t), & t^l \le t < \theta - \varphi \\ Ax_c(t) + Bu_f^b + Da_c(t), & t \ge \theta - \varphi \end{cases} \qquad (13)$$

When $t^l \le t < \theta - \varphi$, $Kx_l(t-\theta)$ can be regarded as an external disturbance, the Eq. (13) can be solved as an ODE. Because the input in the Eq. (13) for $t \ge \theta - \varphi$ is a constant value $u_f^b$, this part can also be solved as an ODE. Then the solution of the Eq. (13) is defined as:

$$x_c(t) = \begin{cases} \underbrace{e^{A(t-t^l)}x_c(t^l) + \int_{t^l}^t e^{A(t-\varepsilon)}BKx_l(\varepsilon-\theta)\,d\varepsilon}_{\text{free response}} + \underbrace{\int_{t^l}^t e^{A(t-\varepsilon)}Da_c(\varepsilon)\,d\varepsilon}_{\text{forced response}}, & t^l \le t < \theta - \varphi \\ \underbrace{e^{A(t-\theta+\varphi)}x_c(\theta) + \int_{\theta-\varphi}^t e^{A(t-\varepsilon)}Bu_f^b\,d\varepsilon}_{\text{free response}} + \underbrace{\int_{\theta-\varphi}^t e^{A(t-\varepsilon)}Da_c(\varepsilon)\,d\varepsilon}_{\text{forced response}}, & t \ge \theta - \varphi \end{cases} \qquad (14)$$

where $x_l(t)$ can be calculated by solving the Eq. (6) based on the general solution of the DDE:

$$x_l(t) = \underbrace{\sum_{k=-\infty}^{+\infty} e^{S_k(t-t^l)} C_k^I}_{\text{free response}} + \underbrace{\int_{t^l}^{t} \sum_{k=-\infty}^{+\infty} e^{S_k(t-\varepsilon)} C_k^N D a_l(\varepsilon) \, d\varepsilon}_{\text{forced response}}, \tag{15}$$

Note that, Eq. (14) gives the deterministic solution given a parameter set $\pi_i = [k_{s,i}, k_{v,i}, k_{a,i}, T_{L,i}, \tau_i, l_i]$. In the real-world these parameters are usually stochastic as suggested by Jiang et al., (2024) and Zhou et al., (2022). Hence, we further consider the empirically calibrated stochastic distribution ACC parameter $\Pi$, following $\Pi = \bigcup_i^M \pi_i$, and $M$ is the total number of parameter sets composite of the joint distribution. By Eq. (14), we can readily define the behavior embedded high-fidelity time to collision $t_{c,i}$ as:

$$hx_c(t_{c,i}|\pi_i) + s_f^*(t) - l' = 0 \tag{16}$$

where $l'$ represents the length of the vehicle, $h = [1 \quad 0 \quad 0]$, $x_c(t_{c,i}|\pi_i)$ is calculated by Eq. (12). Consider the physically meaning, we let $t_{c,i}^*$ as the earliest and positive solution of Eq. (13), otherwise, $t_{c,i}^* = \infty$. By that, we can analyze the safety in a stochastic fashion by compute the expectation of inverse of $t_{c,i}^*$, $E[t_{c,i}^{*,-1}]$ is calculated as following:

$$E[t_{c,i}^{*,-1}] = \frac{\sum_{i=1}^{M} t_{c,i}^{*,-1}}{M} \tag{17}$$

as well as the corresponding cumulative distribution function:

$$C(\gamma) = \frac{\sum_{i=1}^{M} \mathbb{I}_i(\gamma)}{M} \tag{18}$$

where, $\mathbb{I}_i(\gamma)$ is the indication function that

$$\mathbb{I}_i(\gamma) = \begin{cases} 1, & \text{if } t_{c,i}^{*,-1} \leq \gamma \\ 0, & \text{otherwise} \end{cases} \tag{19}$$

Based on the theoretical derivation, if there is no sensing delay, the state evolution of the ACC system depends on the initial state of the cut-in vehicle, the system characteristics, and the external disturbance caused by the acceleration change of the cut-in vehicle. However, if sensing delay is involved, as suggested by the solutions of the DDE in time domains, the initial state of the system includes not only the initial state of the cut-in vehicle but also the state of the original leading vehicle. This represents when there is a sensing delay, the state evolution of the cut-in vehicle and following vehicle is influenced by both the initial state of the cut-in vehicle and state of the original leading vehicle. Additionally, the sensing delay changes the inherent characteristics of the system, with the matrix representing the system's characteristics shifting from $A_f$ to $S_k$. This indicates that sensing delay can impact the stability of the ACC system. According to Eq. (14), the sensing anticipation can eliminate the delay time of the following vehicle's response to the cut-in maneuver. When the sensing anticipation length is equal to the sensing delay length, the following vehicle can response to the vehicle cut-in immediately. When the sensing anticipation length is larger than the sensing delay length, the following vehicle can adjust its acceleration according to the cut-in vehicle before the cut-in occurs. A behavior embedded high-fidelitous SSM is derived from the general solution of the DDE, providing robust and reliable safety assessments of the effects of sensing delays and anticipation on the ACC safety in cut-in scenarios. Jointly calibrated control parameters based on real world commercial ACC datasets are utilized to involve the stochastic of driving behavior.

## 3. Analysis and results

### 3.1. Scenario based analysis

Based on jointly calibrated control parameters, numerical experiments are conducted in this section to investigate the effects of sensing delays and anticipatory sensing on ACC safety in cut-in scenarios. Behavior embedded high-fidelitous SSM are calculated to provide quantitative insights on these effects. Based on theoretical derivations, the vehicle state evolution of sensing delay and anticipation-embedded ACC systems under cut-in scenarios is influenced by the state of the original leading vehicle, the initial state of the cut-in vehicle, subsequent maneuvers of the cut-in vehicle, and the sensing delay and anticipation length. The numerical experiments in this study are designed to include four parts: (i) analysis of vehicle trajectory and velocity dynamics; (ii) sensitivity analysis for the cut-in vehicle state; (iii) sensitivity analysis for original leading vehicle state; (iv) trend analysis for different sensing delays and anticipation length. This study focuses on commercial ACC systems, where the linear feedback control parameters are sampled from empirical joint distributions of control parameters calibrated using an approximate Bayesian computation approach (Jiang et al., 2024; Zhou et al., 2022). Since there are infinite possibilities of the subsequent cut-in vehicle behaviors after merging in, we consider a severe case that the cut-in vehicle undergoes deceleration followed by acceleration, as illustrated by:

$$a_c(t) = \begin{cases} a_1, t \in (0, t_1] \\ a_2, t \in (t_1, t_2) \end{cases} \quad (20)$$

The simulation time step is denoted as $\Delta t$. The default parameter settings for the numerical experiments are listed in Table 1. This study assumes a level road with no grades or horizontal curves.

Table 1: Default parameter settings for the numerical experiments

| Parameters | Value | Parameters | Value |
|---|---|---|---|
| $a_1$ | $-2 \, m/s^2$ | $v_f(t^l)$ | $20 \, m/s$ |
| $a_2$ | $2 \, m/s^2$ | $l'$ | $3 \, m$ |
| $t_1$ | $2 \, s$ | $t^l$ | $-1 \, s$ |
| $t_2$ | $5 \, s$ | $\Delta t$ | $0.1 \, s$ |

To obtain a preliminary understanding of the impact of sensing delays and anticipatory sensing on ACC safety under cut-in scenarios, this section conducted numerical experiments to demonstrate the trajectory, velocity, and gap dynamics of the vehicles. The control parameters and initial conditions of the vehicles are presented in Table 2. As described by the Eq. (20), the cut-in vehicle has a velocity dip after cut-in and the cut-in occurs at 1 second. The original leading vehicle has constant velocity. As suggested by Wang et al. (2018), we selected sensing delay values ranging from 0.1s to 0.3s (Rajamani, 2011). According to existing research, vehicle trajectory prediction with a prediction horizon of 1 second has relatively high accuracy, with a root mean square error of 0.17 on the HighD vehicle trajectory dataset (Wu et al., 2024). Therefore, this study selected 1 second anticipatory length as an example. In this experiment, three following vehicles with different sensing and anticipation abilities are compared. The baseline following vehicle does not involve either sensing delay or anticipation. The second following vehicle incorporates a sensing delay of 0.3 seconds, while the third vehicle includes both a sensing delay of 0.3 seconds and an anticipation time of 1 second. Fig. 3 illustrates the trajectory, velocity, and gap dynamics of the vehicles.

Table 2: Value setting for numerical experiments

| Parameters | Value | Parameters | Value |
|---|---|---|---|
| $k_s$ | $0.26$ | $p_l(0)$ | $100\ m$ |
| $k_v$ | $0.71$ | $v_l(0)$ | $20\ m/s$ |
| $k_a$ | $-1.31$ | $p_f(0)$ | $50\ m$ |
| $\tau^*$ | $1.18\ s$ | $v_f(0)$ | $20\ m/s$ |
| $l$ | $7.64\ m$ | $p_c(1)$ | $110\ m$ |
| $T_L$ | $0.37\ s$ | $v_c(1)$ | $20\ m/s$ |

As shown in Fig. 3 (b), the sensing delay postpones the following vehicle's response (deceleration) to the cut-in event. In contrast, anticipation enables the following vehicle to decelerate before the cut-in occurs. Thus, in this scenario, sensing delay increases the collision risk of the ACC, whereas anticipation reduces it. Specifically, the baseline following vehicle achieves a minimum gap of 3.43 meters (11.25 feet), the sensing delay reduces this gap to 0.77 meters (2.53 feet), and anticipation increases it to 7.12 meters (23.36 feet). Anticipation not only compensates for the influence of the sensing delay but also eliminates the collision risk posed by aggressive cut-ins. Moreover, the sensing delay increases the oscillation of the following vehicle, while anticipation reduces it. Increased oscillation can have several negative impacts, such as a reduction in traffic capacity and increased fuel consumption. Finally, sensing delay and anticipation do not affect the stable state of the vehicles under the cut-in scenario. Specifically, the equilibrium spacing and velocity between the following vehicle and the cut-in vehicle remain constant when either sensing delay or anticipation is involved, which is shown by Fig. 3 (a).

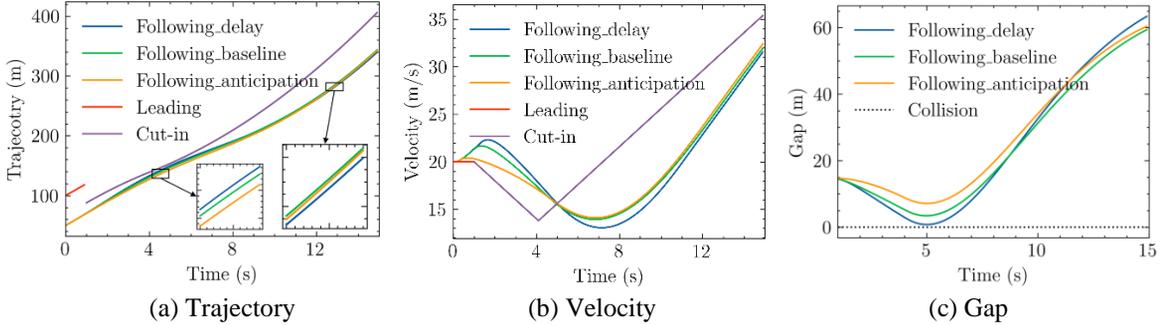

(a) Trajectory  (b) Velocity  (c) Gap

Fig. 3. Illustration of the ACC under the cut-in scenario considering sensing delay and anticipation.

*3.2. Stability analysis*

Based on the theoretical derivations and numerical experiments discussed in the previous section, sensing delay has a substantial impact on the inherent characteristics of ACC, particularly its stability. If the ACC becomes unstable, its state will grow unbounded in response to an initial disturbance, inevitably leading to a collision. This section first validates the influence of sensing delay on ACC stability in a cut-in scenario. The stability of the ACC system without sensing delay is assessed by examining the eigenvalues of the matrix $A + BK$. In this study, 334 sets of control parameters are used. For all parameter sets without sensing delay, the eigenvalues have negative real parts, confirming that the system is asymptotically stable, as expected for a linear time-invariant (LTI) system. To assess the impact of sensing delay on ACC stability, the state evolution of two parameter sets—one stable and one unstable under sensing delay—is analyzed. Fig. 4 presents the dynamics of the spacing deviation under various sensing delay durations. The results are computed numerically using MATLAB. For the first parameter set, the ACC system remains stable with a sensing delay of 0.1 seconds but becomes unstable with delays of 0.2 and 0.3 seconds. Additionally, as the sensing delay increases, the spacing deviation grows more rapidly. In contrast, the second parameter set remains stable across all tested sensing delay durations. However, as shown in Fig. 4, the oscillation amplitude increases with the sensing delay duration, indicating that the effect of sensing delay becomes more pronounced as the delay duration extends. These findings are further supported by the analysis of additional parameter sets, though the corresponding results are not included in this paper due to space limitations.

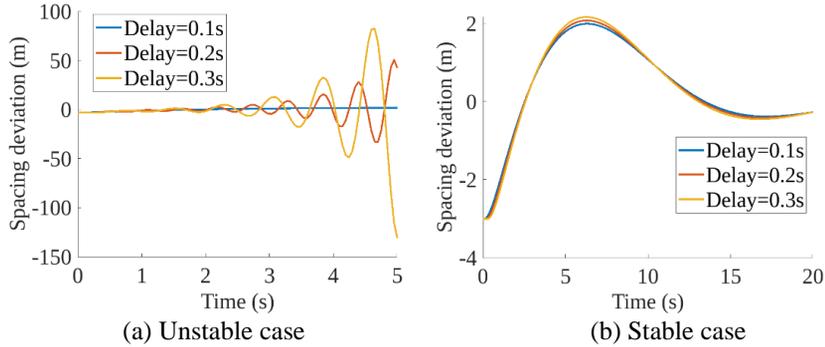

Fig. 4. Spacing deviation ($\Delta s_c(t)$) dynamics for the unstable (a) and stable (b) control parameters.

This study investigates the stability of the control parameter sets under a maximum sensing delay of 0.3 seconds, which is a typical upper limit suggested by existing research (Rajamani, 2011). As concluded in the theoretical derivation section, stability is evaluated by analyzing the eigenvalues of the matrix $S_k$ for branches $k = 0, -1, 1$. If any eigenvalue of $S_k$ has a positive real part, the system is unstable. In such cases, an initial disturbance will cause $x_c(t)$ to grow exponentially, without bound, as time approaches infinity. Conversely, if all the eigenvalues of $S_k$ have negative real parts, the system is asymptotically stable, meaning $x_c(t)$ will tend to zero as time approaches infinity. Although no theoretical proof is provided, all example parameters exhibit this behavior. Experimental results demonstrate that 208 parameter sets are asymptotically stable under a sensing delay of 0.3 seconds. Fig. 5 demonstrates the distribution of the stable parameters.

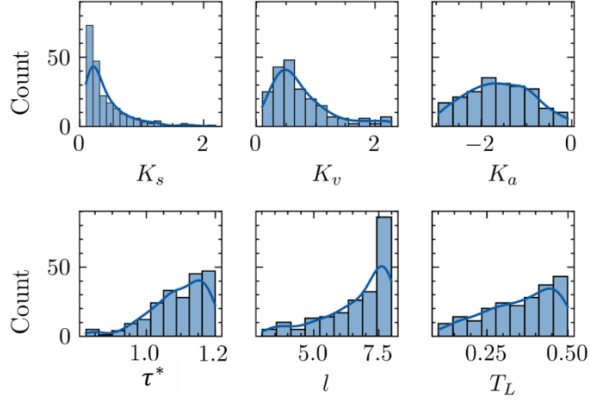

Fig. 5. Distribution of the stable parameters.

*3.3. Sensitivity analysis*

Although the stability of the ACC with sensing delay is validated, the stable ACC does not assure safety when the sensing delay is involved. Thus, this section further investigates the influence of the sensing delay and anticipation on ACC safety under stable control parameters. Based on the theoretical derivation, the vehicle state evolution under cut-in scenarios is dependent on the initial conditions of the cut-in vehicle, the original leading vehicle's condition, and the subsequent maneuvers of the cut-in vehicle. A sensitivity analysis for initial cut-in conditions is first conducted. The range of the initial spacing deviation and velocity deviation between the follower and cut-in vehicle are set as $\Delta s_c(t^l) = [-5m, 0m]$, $\Delta v_c(t^l) = [-5m/s, 0m/s]$ with a resolution of 0.5 m (or m/s). The initial state between the follower and original leading vehicle is set as $\Delta s_l(t^l) = 5m$, $\Delta v_l(t^l) = 5m/s$. $\Delta s_c(t^l)$ and $\Delta v_c(t^l)$ with smaller values represents more aggressive cut-in scenarios. Fifty stable control parameter sets are utilized for the stochastic analysis, vehicle collision is identified for each parameter set, and the probability of collision is obtained after aggregation.

The results indicate that in the absence of sensing delay, all initial cut-in conditions are consistently safe. As illustrated in Fig. 6, the introduction of sensing delay increases the likelihood of collisions due to the delayed deceleration response of the following vehicle to the cut-in event. Specifically, the mean collision probability escalates from zero to 0.1 when a 0.3-second sensing delay is introduced, without the incorporation of anticipation. Fig. 6 (b) and (c) demonstrate that anticipation significantly enhances traffic safety. Anticipation allows the following vehicle to respond to the cut-in vehicle earlier, reducing the mean collision probability from 0.1 to 0.01 as the anticipation time extends from 0 to 0.6 seconds. When the anticipation time reaches 0.6 seconds, the collision probability remains low even under very aggressive cut-in conditions, with all other cut-in scenarios being deemed safe. Notably, when the anticipation time equals the sensing delay at 0.3 seconds, the following vehicle can respond to the cut-in event without any delay. However, as shown in Fig. 6 (b), the risk of collision is not eliminated. This can be attributed to the fact that while the response delay is mitigated, the study assumes that only the anticipation of the cut-in vehicle is considered, leaving the sensing delay related to the original leading vehicle unaddressed, as described by Eq. (14).

Fig. 6 also reveals that when sensing delay is accounted for, more aggressive cut-in scenarios significantly increase the collision risk compared to conservative cut-in scenarios. For instance, when $\Delta s_c(t^l) = -5m$ and

$\Delta v_c(t^l) = -5m/s$, the collision risk surges to 0.8, whereas when $\Delta s_c(t^l) = -3m$ and $\Delta v_c(t^l) = -3m/s$, the risk only rises to 0.02. Additionally, when $\Delta v_c(t^l) > -2m/s$, the cut-in remains safe. These findings suggest that while sensing delay has a minor impact on conservative cut-in scenarios, it can substantially increase the collision risk in more aggressive cut-in scenarios. To further elucidate the relationship between spacing, velocity deviation, and the impact of sensing delay on collision risk, Fig. 7 illustrates the correlation between collision probability and $\Delta s_c$ and $\Delta v_c$. As depicted in Fig. 7, collision probability increases exponentially with decreasing velocity deviation, and linearly with decreasing spacing deviation. This indicates that the impact of sensing delay is more sensitive to velocity deviation, with its effect growing exponentially as the velocity deviation between the cut-in and following vehicles increases. Moreover, this phenomenon highlights that the more pronounced impact of sensing delay on aggressive cut-in scenarios is primarily driven by the velocity deviation between the cut-in and following vehicles.

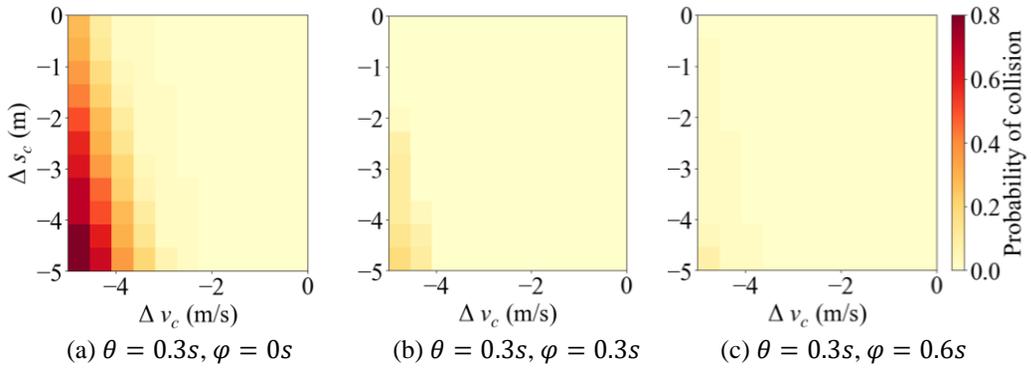

Fig. 6: Probability of collision under distinct cut-in conditions.

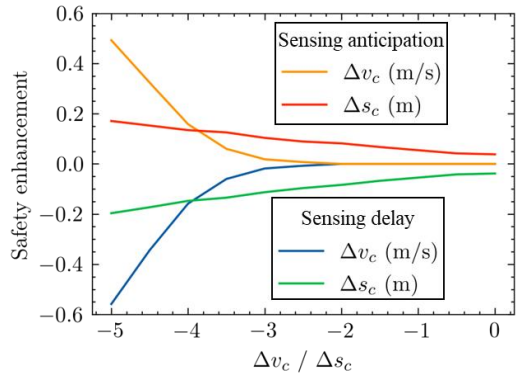

Fig. 7: Safety enhancement: 0.3-second sensing delay/anticipation.

To assess the influence of sensing delay and anticipation under varying states of the original leading vehicle, a sensitivity analysis was conducted focusing on different initial conditions between the follower and original leading vehicle. The initial cut-in condition was set at $\Delta s_c(t^l) = -5m$ and $\Delta v_c(t^l) = -5m/s$, representing a high-risk scenario. The initial state range between the follower and original leading vehicle was defined as

$\Delta s_l(t^l) = [0m, 5m]$ and $\Delta v_l(t^l) = [0m/s, 5m/s]$. In the absence of sensing delay, no collisions were observed across all original leading vehicle conditions. As depicted in Fig. 8, larger initial values of the original leading vehicle's state pose greater risks in cut-in scenarios. This is because, prior to the cut-in, the following vehicle is tracking the original leading vehicle, and larger initial values of the original leading vehicle cause the following vehicle to accelerate, thereby increasing the collision risk between the cut-in and following vehicles. Fig. 8 also illustrates that the impact of sensing delay is more pronounced in cases with higher $\Delta s_l(t^l)$ and $\Delta v_l(t^l)$ values, as these represent more aggressive cut-in scenarios. Furthermore, as shown in Fig. 9, the impact of sensing delay grows exponentially with increasing velocity deviation, consistent with the findings from the sensitivity analysis of the cut-in vehicle's condition. Therefore, it is crucial to moderate the velocity deviations between the following vehicle and both the original leading and cut-in vehicles to mitigate the impact of sensing delay on ACC safety in cut-in scenarios. Fig. 8 (b) and (c) illustrate the influence of anticipation. A 0.3-second anticipation reduces the average probability of collision from 0.21 to 0.02, while a 0.6-second anticipation brings it below 0.01. For aggressive cut-ins, where $\Delta s_l(t^l) = 5m$ and $\Delta v_l(t^l) = 5m/s$, a 0.3-second anticipation decreases the collision probability by 80%, and a 0.6-second anticipation results in a 91% reduction. The effect of anticipation increases exponentially as velocity deviation decreases, and linearly as spacing deviation decreases. The possible reason is that the safety-enhancing effect of anticipation is more pronounced in more hazardous cut-in scenarios. Fig. 10 illustrates the trajectories of vehicles in a high-risk cut-in scenario ($\Delta s_l(t^l) = \Delta v_l(t^l) = 5$ m (m/s) and $\Delta s_c(t^l) = \Delta v_c(t^l) = -5$ m (m/s)), where a stable control parameter set is applied. When a 0.3-second sensing delay is present, the following vehicle collides with the cut-in vehicle at approximately 5 seconds. In contrast, without sensing delay, the gap between the following and cut-in vehicles reaches 4.82 meters (15.81 feet) and incorporating a 0.6-second anticipation improves the gap to 6.62 meters (21.72 feet), effectively avoiding a collision.

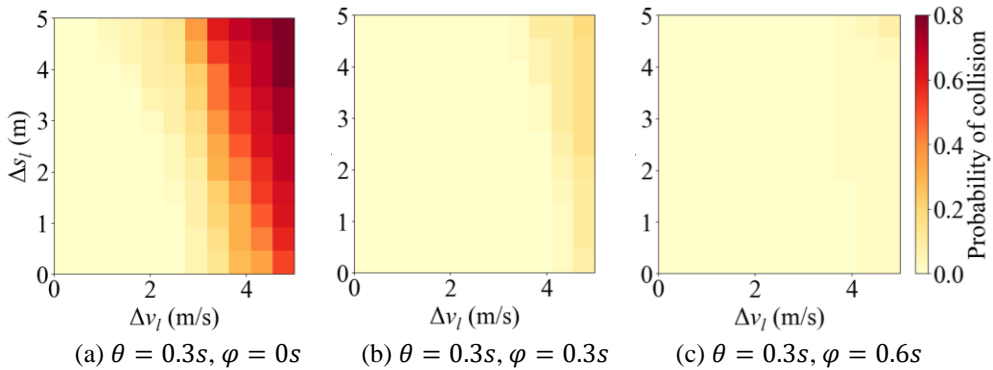

(a) $\theta = 0.3s, \varphi = 0s$    (b) $\theta = 0.3s, \varphi = 0.3s$    (c) $\theta = 0.3s, \varphi = 0.6s$

Fig. 8: Probability of collision under distinct original leading vehicle conditions.

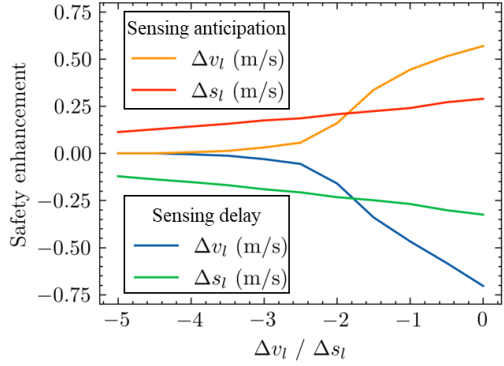

Fig. 9: Safety enhancement: 0.3-second sensing delay/anticipation.

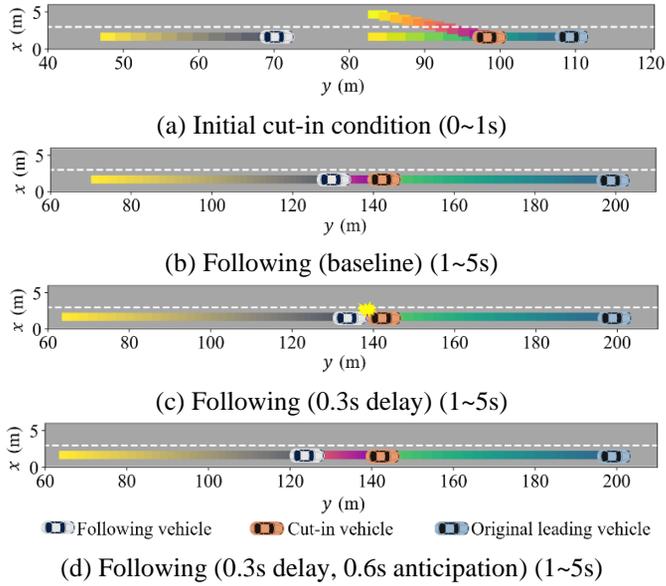

(a) Initial cut-in condition (0~1s)

(b) Following (baseline) (1~5s)

(c) Following (0.3s delay) (1~5s)

(d) Following (0.3s delay, 0.6s anticipation) (1~5s)

Fig. 10: Illustration of vehicles' trajectory. (Note: The colour range from light to dark represents the progression of time from past to recent)

*3.4. Analysis for various sensing delay and anticipation length*

To further investigate the impact of sensing delay and anticipation, a range of values for both parameters were tested in this experiment. The distribution of the inverse time to collision (TTC) was analyzed using two hundred parameter sets. In this study, inverse TTC is defined as the inverse of the $t_{c,i}^*$, which is calculated by Eq. (16). Based on insights from existing studies, the sensing delay range was set between 0 and 0.3 seconds, while the anticipation range was set between 0 and 2 seconds, with both parameters sampled at a resolution of 0.1 seconds. Additionally, theoretical derivations indicate that the state evolution of the ACC system is influenced by the cut-in vehicle's maneuvers. Therefore, this section examines the impact of various cut-in vehicle maneuvers on the ACC system. This study also considers the accuracy of predicting cut-in behavior, as incorrect predictions of false cut-in events do not increase collision risk. Specifically, the analysis focuses on true cut-in behavior prediction accuracy, with an interaction-aware cut-in prediction algorithm proposed by Zhu et al., (2022) achieving a high prediction accuracy of 99.7%. This prediction accuracy is incorporated into the analysis of anticipation's impact.

Fig. 11 illustrates the collision probability across four scenarios with distinct cut-in vehicle maneuvers and conditions, with the original leading vehicle's initial state set at $\Delta s_l(t^l) = 5m$ and $\Delta v_l(t^l) = 5m/s$. As shown in Fig. 11, a further dip in the cut-in vehicle's velocity significantly increases the collision risk. The influence of sensing delay is exacerbated by this velocity dip, corroborating the conclusion from the previous section that sensing delay has a more pronounced impact in more hazardous cut-in scenarios. For instance, Fig. 11 (a) and (c) reveal that a 0.3-second sensing delay increases the collision probability to 0.82 when there is a velocity dip, compared to just 0.19 in the absence of such a dip. Fig. 11 also shows that collision probability increases with longer sensing delays, while it decreases with greater anticipation. A longer anticipation time is required to mitigate collision risk as sensing delay increases. For example, as shown in Fig. 11 (a), when the sensing delay is 0.1 seconds and 0.2 seconds, anticipation times of 0.4 seconds and 0.6 seconds, respectively, are necessary to eliminate collision risk. Fig. 12 depicts the distribution of inverse TTC values as calculated by Eq. (18). As illustrated in Fig. 12, sensing delay has a minor effect on the peak location of the distribution. However, it increases the magnitude of inverse TTC values greater than zero and extends the tail of the distribution. Conversely, anticipation has the opposite effect, reducing both the magnitude of the distribution for values greater than zero and shortening the distribution's tail.

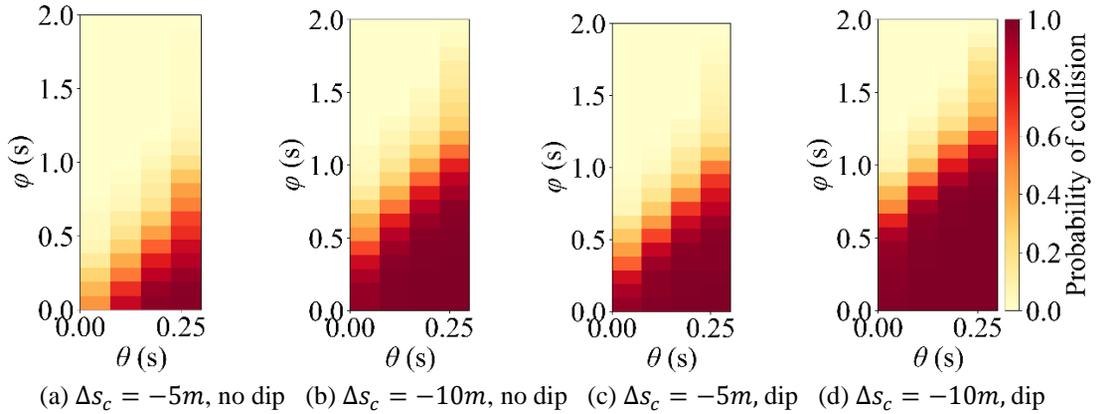

(a) $\Delta s_c = -5m$, no dip  (b) $\Delta s_c = -10m$, no dip  (c) $\Delta s_c = -5m$, dip  (d) $\Delta s_c = -10m$, dip

Fig. 11: Probability of collision under distinct sensing delays and anticipation duration.

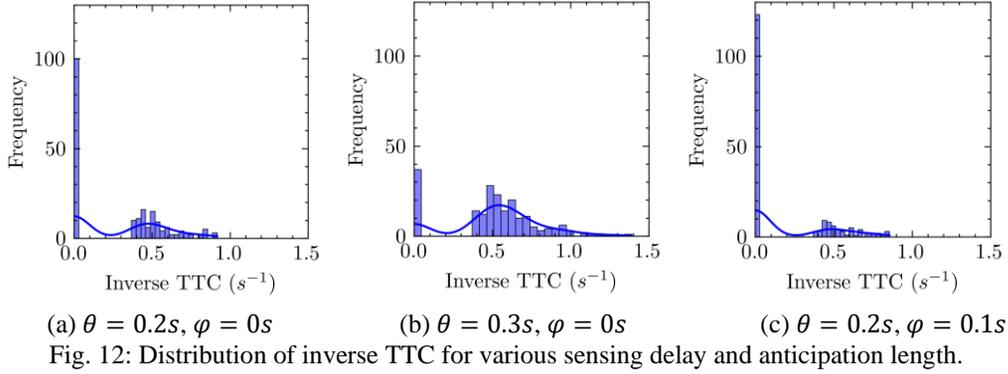

Fig. 12: Distribution of inverse TTC for various sensing delay and anticipation length.

## 4. Conclusions

This study presents an analytical solution for the vehicle state evolution in ACC systems under cut-in scenarios as well as a stochastic behavior embedded high-fidelitous SSM, incorporating the effects of sensing delays and anticipation. Using the Lambert W function, the analytical solution models the system's linear feedback control with embedded sensing delays, based on the general solution of delay differential equations. The free response is driven by the initial conditions of the cut-in and original leading vehicles, while the forced response is shaped by subsequent cut-in maneuvers and sensing delays. Anticipation is accounted for by adjusting the timing of the following vehicle's response to the cut-in event. The theoretical analysis highlights that ACC safety in cut-in scenarios is influenced by various factors: original leading vehicle conditions, the initial state of the cut-in vehicle, its maneuvers, sensing delays, and anticipation. Importantly, sensing delays are shown to affect ACC system stability, leading to increased safety risks in the system when disturbed. On the contrary, anticipation plays a key role in enhancing safety by mitigating the risks.

Numerical experiments were conducted to quantify these effects, revealing that sensing delays of 0.1 to 0.3 seconds can increase the oscillations in the ACC system and reduce stability. Specifically, with a sensing delay of 0.3 seconds, the collision risk increased to 80% in aggressive cut-in scenarios. The risk was further amplified when the cut-in vehicle's speed was 10-20% slower than the following vehicle's. Anticipation, however, significantly mitigates collision risk, with a 0.6-second anticipation reducing the likelihood of collisions by up to 91% in high-risk scenarios. As anticipation extends to 2 seconds, safety is maintained even under severe conditions, effectively countering the adverse effects of sensing delays. These findings provide a quantitative foundation for understanding how sensing delays and anticipation jointly impact ACC safety.

Overall, this study underscores the importance of optimizing both sensing delays and anticipation mechanisms in ACC systems to enhance safety and performance in cut-in scenarios. Further research could explore how these factors interact with different road conditions and vehicle dynamics to develop even more robust ACC solutions.

## Acknowledgements

This research is supported by the Federal Highway Administration (FHWA) Exploratory Advanced Research (693JJ323C000010). The results do not reflect FHWA's opinions.